\documentclass[]{aastex}

\usepackage{emulateapj5}
\usepackage{onecolfloat}
\usepackage{graphicx} 
\usepackage{fancyheadings} 
\usepackage{ulem}
\usepackage{rotating}
\usepackage{lscape}

\newcommand{\kms}{km~s$^{-1}$ }
\newcommand{\cm}[1]{\, {\rm cm^{#1}}}
\newcommand{\mkms}{{\rm \; km\;s^{-1}}}
\newcommand{\delv}{\Delta v}

\newcommand{\lya}{Ly$\alpha$}

\newcommand{\N}[1]{{N({\rm #1})}}
\newcommand{\sci}[1]{{\rm \; \times \; 10^{#1}}}

\newcommand{\mnhi}{N_{\rm HI}}

\newcommand{\nhi}{$N_{\rm HI}$}

\def\nhi{$N_{\rm HI}$}

\begin{document}

\twocolumn[%
\submitted{Accepted to ApJ, June 19 2006}

\title{On the Perils of Curve-of-Growth Analysis:
Systematic Abundance Underestimates for the Gas in
Gamma-Ray Burst Host Galaxies}

\author{Jason X. Prochaska\altaffilmark{1}}

\begin{abstract}
We examine the practice of deriving interstellar medium (ISM) 
abundances from low-resolution spectroscopy of GRB afterglows.
We argue that the multi-ion single-component curve-of-growth
analysis technique systematically underestimates the column
densities of the metal-line profiles commonly observed
for GRB.  This systematic underestimate is accentuated by the fact
that many GRB line-profiles (e.g.\ GRB~050730, GRB~050820, GRB~051111)
are comprised of `clouds' with a bi-modal 
distribution of column density.  Such line-profiles may be 
characteristic of a sightline which penetrates both a high
density star-forming region and more distant, ambient ISM material.
Our analysis suggests that the majority of abundances reported
in the literature are systematically underestimates and that the 
reported errors are frequently over-optimistic.
Further, we demonstrate that one cannot even report relative
abundances with confidence.
The implications are profound for our current understanding
on the metallicity, dust-to-gas ratio, and chemical abundances
of the ISM in GRB host galaxies.  For example, we argue that
all but a few sightlines allow for the gas to have at least
solar metallicity.  Finally, we suggests new approaches 
for constraining the abundances.

\end{abstract}
]

\altaffiltext{1}{Department of Astronomy and Astrophysics, 
UCO/Lick Observatory;
University of California, 1156 High Street, 
Santa Cruz, CA 95064; xavier@ucolick.org}

\pagestyle{fancyplain}
\lhead[\fancyplain{}{\thepage}]{\fancyplain{}{PROCHASKA}}
\rhead[\fancyplain{}{On the Perils of Curve-of-Growth Analysis}]{\fancyplain{}{\thepage}}
\setlength{\headrulewidth=0pt}
\cfoot{}

\section{Introduction}

With the launch of the {\it Swift} satellite
\citep{gcg+04}, the rate of GRB detections has increased
by more than an order of magnitude \citep{gehrels06}.  In turn, 
one has seen roughly the same increase in the detection of 
bright afterglows.  
Though a few high-resolution spectra were obtained 
pre-Swift \citep{cgh+03,fdl+05}, Swift has enabled the nearly routine
acquisition of high signal-to-noise ratio (SNR), high-resolution spectra 
\citep{cpb+05,pro_finegrb06}.  In addition,  there are now 
over a dozen low-resolution
observations from pre-Swift GRB \citep[e.g.][]{kdo+99,bsc+03}
and the first year of Swift operation
\citep[e.g.][]{foley06,bpck+05,fynbo060206}. 
Observers are now
progressing toward the examination of
distributions of gas properties in the interstellar medium (ISM) 
of GRB host galaxies.

Because of the apparent faintness of many GRB afterglows and also instrument
availability, the majority of GRB spectra are
acquired with low-resolution spectrometers (FWHM$>1$\AA) and
at moderate to poor SNR ($<50$ per resolution element).  In these cases,
abundance analysis must be performed on the equivalent width
measurements of unresolved metal-line
transitions.
To date, the standard practice has been to perform a multi-ion 
single-component curve-of-growth (MISC-COG) analysis
\citep[e.g.][]{sff03} or
single-component profile fits to unresolved data 
\citep[e.g.][]{sf04,watson050401} which is fundamentally the same analysis.
In this analysis, one derives an effective Doppler parameter $b_{eff}$
which principally characterizes the velocity width of strong, 
heavily saturated transitions.  The column density is then 
constrained with weaker, yet still potentially saturated transitions.
In contrast, the \ion{H}{1} column density
can be determined from fits to the damping wings of the \lya\ profile
even with low-resolution data \citep[e.g.][]{phw05}.

The curve-of-growth technique dates far back \cite[e.g.][]{unsold30,wilson39} 
and was primarily introduced to perform abundance analysis on lower
spectral resolution data.
The literature on the pitfalls of MISC-COG analysis
is extensive \citep{nh73,cru75,spitzer75}.  These include systematic differences
in the results for ions of differing ionization state 
\citep[e.g.\ \ion{Mg}{1} vs.\ \ion{Fe}{2};][]{spitzer75}, `hidden' saturation
due to significant variations in the Doppler parameter of overlapping
components \citep{nh73}, and the dangers of mixing 
refractory and non-refractory elements 
\citep{jss86}.  The Galactic ISM community is
well aware of these issues; it has taken every effort to obtain
high resolution spectroscopy and when limited to low resolution data,
very cautiously interpret the observations.  
In short, few of the problems discussed in this paper are unique
to GRB research, but we note that the discussion may translate
to other abundance analyses with large, modern datasets of low
resolution spectroscopy \citep[e.g.][]{sga+04,trn+05}.

\cite{jenkins86}  examined the accuracy of 
a single-component COG analysis on multi-component line-profiles for a range
of reasonable distributions of column density and Doppler parameter.
He showed that the COG results typically give results accurate to
within 20\% (i.e.\ $<0.2$\,dex)
of the correct column density provided the $N$ and $b$
distributions of the individual components are well-behaved and so long
as the peak optical depth $\tau_0 < 5$.  
Jenkins also emphasized,
however, that the results would likely significantly underestimate
the column densities if the $N$ or $b$ values were bi-modally distributed
\citep[see also][]{nh73}.  We will find that this is frequently the
case for GRB line-profiles.

In this paper, we will examine the main issues related to the ISM abundance
analysis of GRB host galaxies as derived from low-resolution
spectroscopy, i.e.\ equivalent width analysis.
We begin with an analysis of the GRB~051111 sightline where 
Keck/HIRES observations permit one to directly test the MISC-COG technique 
($\S$~\ref{sec:051111}).
The discouraging results are enlightening, namely the COG analysis
systematically underestimates the column density measurements.  
We consider an additional example from the literature (GRB~020813) 
and raise similar concerns ($\S$~\ref{sec:020813}).
Finally, we offer a set of guidelines to
consider when analyzing low-resolution GRB spectra ($\S$~\ref{sec:proc}). 
In a future paper \citep{pro_dataI06}, we will apply these guidelines
to our datasets and previously published results with the intent
of providing a uniform, database of measurements with realistic
error estimates.

\begin{figure}[ht]
\begin{center}
\includegraphics[width=3.5in]{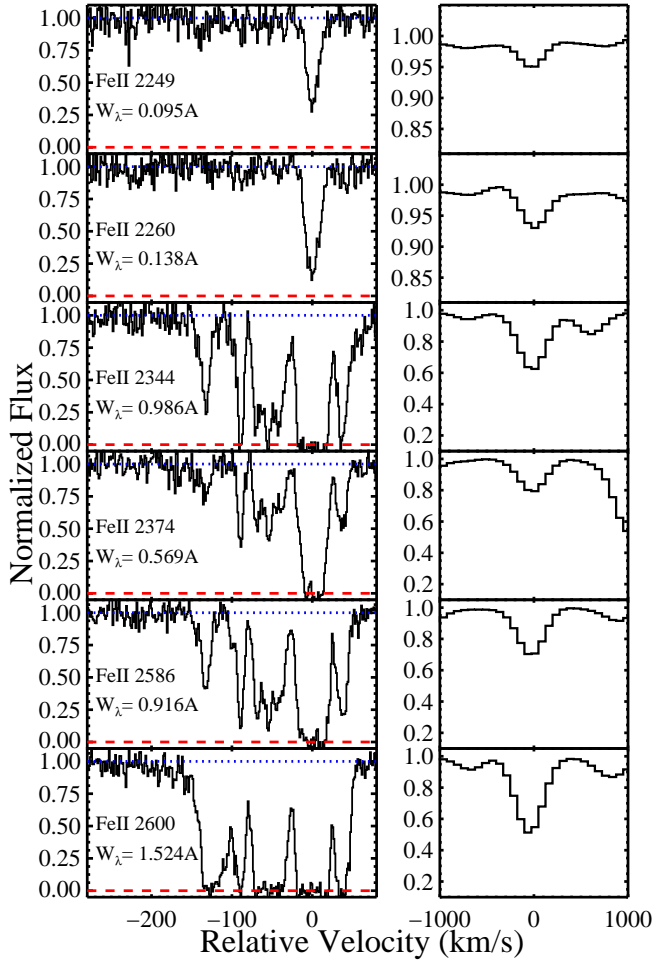}
\caption{\ion{Fe}{2} profiles from the ISM foreground to
GRB~051111 in the original data (LHS) and smoothed to lower
resolution (RHS; FWHM~$\approx 5$\AA) 
to mimic the line-profiles frequently observed.  
The velocity $v=0$ corresponds to $z=1.54948$.
The rest equivalent width values labeled on the Figure correspond to the
full line-profile.
}
\label{fig:fe051111}
\end{center}
\end{figure}

\section{GRB~051111: A Test Case}
\label{sec:051111}

To illustrate the key issues related to deriving abundances from
equivalent width measurements, we will examine a GRB sightline 
where the column densities are well-constrained, i.e., 
an afterglow with high quality, high resolution observations.
The Keck observatory staff acquired HIRES \citep{vogt94}
observations of the bright afterglow of the $z \approx 1.55$
GRB~051111 \citep{gcn4255} at high spectral resolution 
FWHM~$\approx 6 \mkms$ and moderate signal-to-noise ratio 
($\approx 7$ per 1.3\kms\ pixel) and released the data to the 
public\footnote{http://www.graasp.org}.   
The data reveal
the detection of over 50 transitions \citep{pro_dataI06} including
a wealth of fine-structure levels \citep{pro_finegrb06}.  
Our treatment begins with a single-component COG analysis of
the Fe$^+$ ion and then proceeds to a MISC-COG analysis of other
ions in the ISM surrounding GRB~051111.

\begin{table}[ht]\footnotesize
\begin{center}
\caption{{\sc EQUIVALENT WIDTH MEASUREMENTS FOR GRB~051111\label{tab:ew}}}
\begin{tabular}{lccccrr}
\tableline
\tableline
Ion & J$^a$ & $\lambda$ & $\log f$
& $W_\lambda^b$ & $N_{adopt}^c$ & $\tau_{0}^d$ \\ 
& & (\AA) & & (m\AA) & ($\cm{-2}$) \\
\tableline
\ion{Ni}{2}\\
& & 1709.604 & -1.4895 &$  54.2 \pm  7.2$&$ 13.97 $&$  1.1$\\
& & 1741.553 & -1.3696 &$  76.6 \pm  6.8$&$$&$  1.5$\\
& & 1751.916 & -1.5575 &$  48.5 \pm  6.8$&$$&$  1.0$\\
\ion{Mg}{1}\\
& & 1747.794 & -2.0419 &$  72.3 \pm  6.1$&$ 14.68 $&$  1.6$\\
& & 1827.935 & -1.6216 &$ 133.3 \pm  5.9$&$$&$  4.5$\\
& & 2026.477 & -0.9508 &$ 279.7 \pm  4.3$&$$&$ 23.2$\\
\ion{Si}{2}\\
&1/2 & 1808.013 & -2.6603 &$ 338.4 \pm  7.8$&$> 16.14$&$> 11.7$\\
\ion{Zn}{2}\\
& & 2026.136 & -0.3107 &$ 237.9 \pm  4.4$&$> 13.71$&$> 10.9$\\
& & 2062.664 & -0.5918 &$ 176.7 \pm  5.0$&$$&$>  5.8$\\
\ion{Cr}{2}\\
& & 2056.254 & -0.9788 &$ 154.9 \pm  5.1$&$ 13.88 $&$  3.5$\\
& & 2066.161 & -1.2882 &$  88.8 \pm  5.5$&$$&$  1.7$\\
\ion{Fe}{2}\\
&9/2 & 2249.877 & -2.7397 &$ 112.9 \pm  6.8$&$ 15.32 $&$  1.8$\\
&9/2 & 2260.780 & -2.6126 &$ 155.1 \pm  6.5$&$$&$  2.5$\\
&9/2 & 2344.214 & -0.9431 &$ 985.6 \pm  9.5$&$$&$120.0$\\
&9/2 & 2374.461 & -1.5045 &$ 569.0 \pm  7.9$&$$&$ 33.4$\\
&9/2 & 2586.650 & -1.1605 &$ 916.4 \pm  7.9$&$$&$ 80.2$\\
&9/2 & 2600.173 & -0.6216 &$1523.8 \pm  7.8$&$$&$279.0$\\
\ion{Fe}{2}\\
&7/2 & 2333.516 & -1.1601 &$ 205.9 \pm  4.3$&$ 14.01 $&$  3.5$\\
&7/2 & 2396.356 & -0.5414 &$ 405.8 \pm  4.9$&$$&$ 15.0$\\
\ion{Mn}{2}\\
& & 2576.877 & -0.4549 &$ 260.6 \pm  5.1$&$> 13.64$&$>  8.5$\\
& & 2594.499 & -0.5670 &$ 229.5 \pm  4.1$&$$&$>  6.6$\\
& & 2606.462 & -0.7151 &$ 213.2 \pm  5.3$&$$&$>  4.7$\\
\tableline
\end{tabular}
\end{center}
\tablenotetext{a}{J value for ions with excited states.}
\tablenotetext{b}{Rest equivalent width.}
\tablenotetext{c}{Weighted-mean value.}
\tablenotetext{d}{Peak optical depth of the line profile assuming the 
column \\ density from Column~6 and $b=$ 7\kms.}
\end{table}

\subsection{Single-component COG Analysis}

\begin{figure}[ht]
\begin{center}
\includegraphics[width=3.6in]{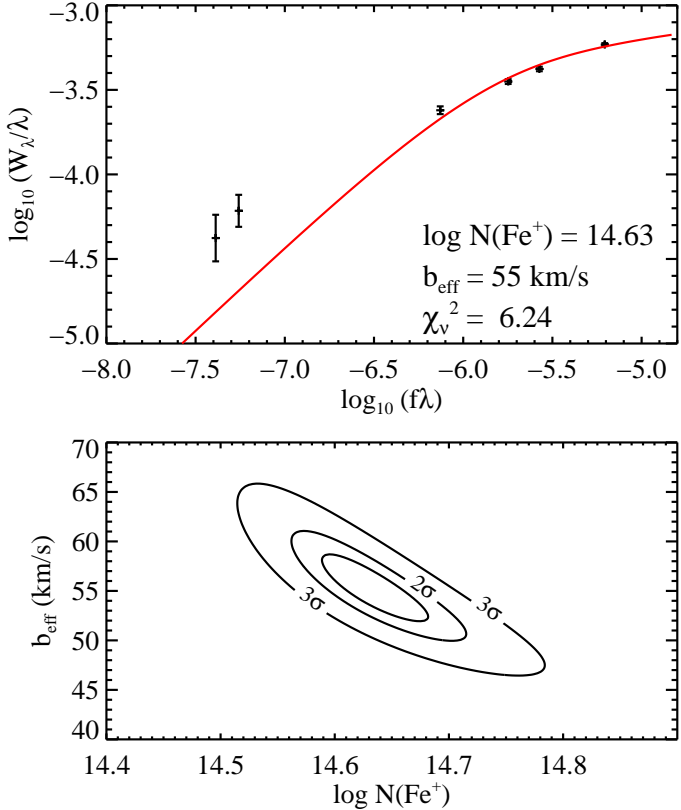}
\caption{Single-component COG analysis of the \ion{Fe}{2} transitions
for GRB~051111 under the assumption that the equivalent width error
$\sigma(W_\lambda) = 30$m\AA\ for each transition.  The upper panel
presents the $W_\lambda$ measurements and the model which minimizes
$\chi^2$.  The lower panel shows $\Delta \chi^2$ contours for the 
$b_{eff}$, $N$ parameter space.  The best-fit column density (and $2\sigma$
error ellipse) severely underestimate the true column density
$N_{true} = 10^{15.24} \cm{-2}$.
}
\label{fig:sngcog051111}
\end{center}
\end{figure}

We will focus first on the \ion{Fe}{2} transitions (Figure~\ref{fig:fe051111})
from the ground-state
($J=9/2$) level and will adopt the Fe$^+$ column
density $\N{Fe^+} = 10^{15.27 \pm 0.01} \cm{-2}$ from \cite{pro_finegrb06}.
We have measured the total rest equivalent width of each
\ion{Fe}{2} transition 
and its associated statistical uncertainty (Table~\ref{tab:ew}).
At the resolution and signal-to-noise of these HIRES data, 
the rest equivalent width statistical uncertainty at 6000\AA\ in one pixel is
$\sigma(W_\lambda) = 1.5$m\AA.  To better match the
low resolution observations, consider an analysis assuming total 
uncertainty $\sigma(W_\lambda) = 30$m\AA.
Figure~\ref{fig:sngcog051111} presents a single-component COG analysis
of the \ion{Fe}{2} transitions which is a two parameter fit to the
data: the column density $N_{best}$ and the effective Doppler
parameter $b_{eff}$.  
The upper panel shows the reduced equivalent widths
$W_\lambda / \lambda$ plotted against the $f\lambda$ product.
The single-component model which minimizes $\chi^2$ is overplotted 
on the observations.  
The model is driven to match the high $W_\lambda$ transitions because
for fixed $\sigma(W_\lambda)$ these measurements have the highest 
relative SNR.  This is a generic result for the analysis of
unresolved line-profiles.
The lower panel presents the $\Delta \chi^2$
contours for the $N, b_{eff}$ parameter space.
The best-fit model (and its $2\sigma$ error ellipse) underestimates
the correct total Fe$^+$ column density by $0.6$\,dex, i.e.\ a factor
of 4.

Why does the single-component COG model systematically underestimate
the Fe$^+$ column density?  The principal reason is that a single-component
line-profile
is only a good representation of the two weakest \ion{Fe}{2} transitions 
(Figure~\ref{fig:fe051111}).  In order to match the observed
equivalent width of the strong \ion{Fe}{2} transitions which are
comprised of many ($>10$) `clouds', a single-component COG analysis
is driven to a large effective Doppler parameter, i.e.\ $b_{eff} > 50\mkms$.
The only physical significance of this large $b_{eff}$
value is that it describes
the total velocity width of the very strongest lines 
($\delv \approx 200 \mkms$).  For this sightline, over $90\%$ of the 
Fe$^+$ ions are associated with the component at $v=0\mkms$ and
confined to $\delv \approx 20 \mkms$. 
Meanwhile, the gas responsible for $>80\%$ of the equivalent width
accounts for less than $10\%$ of the total column density.
It is this bi-modality in component column densities and the inclusion
of highly saturated transitions which leads to such
poor results for the COG analysis.

\begin{figure}[ht]
\begin{center}
\includegraphics[height=3.6in,angle=90]{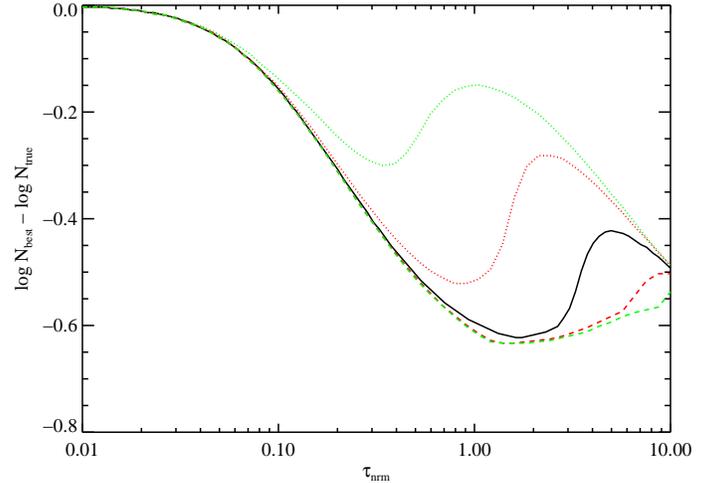}
\caption{Offset between the best-fit column density $N_{best}$
for a single-component COG analysis and the true column density
$N_{true}$.  In this analysis, we have adopted the Fe$^+$ optical
depth profile from the unsaturated regions of the
\ion{Fe}{2}~2260, 2374, and 2586 transitions for GRB~051111.
We then varied the normalization of the optical depth profile $\tau_{nrm}$,
measured the $W_\lambda$ values for each of the transitions in
Figure~\ref{fig:fe051111}, and minimized $\chi^2$ for $N_{best}$
and $b_{eff}$.
The solid curve shows the offsets assuming $\sigma(W_\lambda)=0.03$m\AA\
for each transition.  The dotted curves show the results assuming
$\sigma(W_\lambda)$ is twice and four times smaller for the 
weakest transitions (\ion{Fe}{2}~2249 and 2260).  Finally, the
dashed lines show the results assuming $\sigma(W_\lambda)$
is 2$\times$ and 4$\times$ larger for the weakest transitions.
}
\label{fig:scogvstau}
\end{center}
\end{figure}

The best-fit $N_{best}$ value is sensitive to the
error estimate $\sigma(W_\lambda)$ that we adopt, although only
if we alter the {\it relative} uncertainty.  For example, if we 
set $\sigma(W_\lambda) = 40$m\AA\ for each transition
then we will derive the same results as
in Figure~\ref{fig:sngcog051111}, albeit with a larger error ellipse.
This is because, the relative $\chi^2$ of each line 
is unaltered.  If we instead
we increase $\sigma(W_\lambda)$ for only the strongest transitions, the
best-fit model slowly approaches the correct value.
Figure~\ref{fig:scogvstau} describes these trends.  For this Figure,
we have first derived the optical depth profile of the Fe$^+$ gas
by combining the unsaturated observations of the \ion{Fe}{2} 2260,
2374, and 2586 profiles.  We then varied
the normalization of the optical
depth profile ($\tau_{nrm} = 1$ corresponds to the data as observed),
measured the equivalent widths of the transitions,
and minimized $\chi^2$ to derive $N_{best}$ 
and $b_{eff}$.  The solid curve shows
the results assuming $\sigma(W_\lambda) = 30$m\AA\ for each transition.
At small $\tau_{nrm}$, the transitions are optically thin and one
recovers the correct result independent of the best-fit $b_{eff}$ value.
At $\tau_{nrm} > 1$, the $N_{best}$ value is
systematically lower than $N_{true}$.
The dotted curves in the Figure show the results when we
reduce the uncertainty in the \ion{Fe}{2}~2249,2260
transitions by a factor of 2 and 4 respectively.  This leads to a 
significant improvement for $\tau_{nrm} < 5$ where the lines are
at most modestly saturated.  The dashed lines show the results where
$\sigma(W_\lambda)$ is increased by 2 and 4 times for the weak transitions.
The key conclusions to be drawn from Figures~\ref{fig:sngcog051111}
and \ref{fig:scogvstau} is that the standard single-component COG
analysis may lead to a severe underestimate of the gas column density.

There are three signatures that the single-component
COG analysis is providing an underestimate of the column density.  
First, the effective Doppler parameter
exceeds 50\kms.  Low-ion line-profiles in the 
Galactic ISM \citep[e.g.][]{spitzer95,howk99}, 
damped \lya\ systems \citep[e.g.][]{prochaska96,mirka03}, and
GRB sightlines \citep[e.g.][]{cpb+05} all break into individual
components with Doppler parameters $b<20\mkms$.  This is especially
true for the cloud which corresponds to the peak optical depth of the
line.  On its own, however, the derivation of a large $b_{eff}$
value does not require that the measurement is erroneous \cite{jenkins86}.  But, 
if one adopts $b_{eff} \gg 20\mkms$ then {\it one will calculate
column densities for weaker transitions
in the optically thin limit even if these 
transitions are actually saturated.}
A second signature is that the reduced $\chi_\nu^2$ is large.  
This is likely to be
a universal result for GRB observations, again because the single-component
model is a poor representation of the data.  The key concern here is that
the error estimate that one derives from a $\Delta \chi^2$ analysis
will generally by far too optimistic.
Finally, the most important signature is when the 
reduced equivalent width of the
weakest transitions are significantly under-predicted by the model.
This is the most striking feature in Figure~\ref{fig:sngcog051111}.
We note that this effect is also evident, for example, 
in the Fe$^+$ measurements
for GRB~000926, GRB~010222, and GRB~050505 as analyzed by
\cite{sff03} and \cite{bpck+05}. 

\begin{figure}[ht]
\begin{center}
\includegraphics[height=3.6in,angle=90]{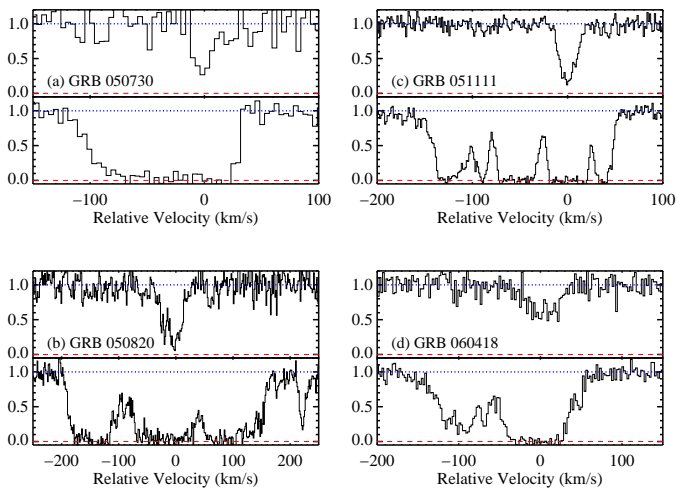}
\caption{Velocity profiles for a weak and strong transition
arising in the ISM of the host galaxies of 
(a) GRB~050730, (b) GRB~050820, (c) GRB~051111, and (d) GRB~060418
\citep{pro_dataI06}.
Note that the majority of column density is associated with a
single, narrow component whereas the equivalent width of the
strongest transitions is dominated by weaker `clouds'.
It is this bi-modal aspect of the column density distribution
which accentuates the failings of the single component COG
analysis.
}
\label{fig:bimodal}
\end{center}
\end{figure}

Before proceeding to the multi-ion, single-component COG analysis,
we wish to comment on our results in light of the analysis by 
\cite{jenkins86}.  
As noted in the Introduction, \cite{jenkins86} warned that 
the single-component COG analysis
would significantly underestimate
the column density in cases where the optical depth distribution
for the individual components is not smoothly distributed
(i.e.\ bi-modal).
Indeed, this is the case for the line-profiles
for GRB~051111 (Figure~\ref{fig:fe051111}).  We emphasize that this
is also the case for the ISM gas associated with GRB~050730 and GRB~050820,
i.e.\ every case where moderate SNR, high-resolution echelle data
exists \citep{cpb+05,pro_dataI06}.  
Again, the total optical depth of the line-profiles 
is dominated by a single, narrow
component and the total equivalent width (for strong transitions)
is dominated by low column density clouds. 
We expect that a bi-modal distribution of $N$ values will be characteristic
of many GRB sightlines\footnote{In passing, we also note that 
a bi-modal column density distribution is also
characteristic of many of the damped \lya\ systems \citep[e.g.][]{pro01}.}: 
the high column density component may 
be related to the surrounding star-forming region and the lower column
density gas may arise from the ambient ISM or even halo of the galaxy.
Figure~\ref{fig:bimodal} compares one weak and one strong transition
from the ISM of the host galaxies of GRB~050730, GRB~050820, GRB~051111,
and GRB~060418 \citep{pro_dataI06}.  It is evident that the majority
of column density is in a single narrow component while the equivalent
width is dominated by multiple, weaker components.  Of course,
additional high-resolution observations of GRB afterglows will help
reveal the typical $N$ and $b$ distributions of the line-profiles.

\begin{figure}[ht]
\begin{center}
\includegraphics[width=3.6in]{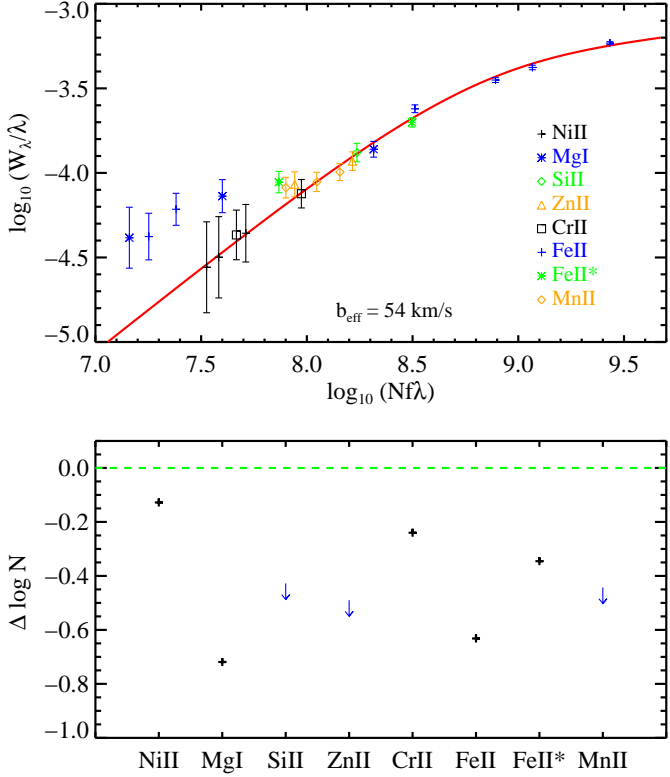}
\caption{The multi-ion single-component curve-of-growth (MISC-COG)
analysis for a series of ions observed in the ISM of GRB~051111.
The upper panel shows the best-fit model and the observations.  
The lower panel shows the offset for the best-fit column density
and the actual value.  The arrows indicate lower limits to the
offsets because the \ion{Si}{2}, \ion{Zn}{2}, and \ion{Mn}{2}
transitions are all saturated at echelle resolution.
}
\label{fig:mscog051111}
\end{center}
\end{figure}

\subsection{Multi-ion Single-component COG Analysis}

Let us now consider a multi-ion single-component COG (MISC-COG)
analysis for the GRB~051111 transitions.  To date, this has been the
standard approach for analyzing low-resolution GRB spectra.
In this analysis, one minimizes the global $\chi^2$ 
for a single $b_{eff}$ value
and unique $N_{best}$ values for each ion considered.
Figure~\ref{fig:mscog051111} shows the results for a number of ions
observed toward GRB~051111.  As before, we have assumed
$\sigma(W_\lambda) = 30$m\AA\ for all of the transitions.
Aside from the weakest \ion{Mg}{1} and \ion{Fe}{2} transitions,
we find the data is reasonably well represented by this model, primarily
because each ion has an additional degree of freedom associated with it.
In fact, one could easily choose to ignore the `discordant' 
\ion{Mg}{1} and \ion{Fe}{2} 
points on account of line-blending and/or an unidentified cosmic-ray
(especially if the observations include only one or two examples).
We emphasize, however, that these data points are severely discrepant
because (i) the analysis includes at least one strong, heavily saturated
transition ($\tau_0 > 5$)
from each ion which has driven $N_{best}$ to a systematically
low value; and (ii) the transitions are optically thin and therefore
highlight that the $N_{best}$ value is too low.

The lower panel of Figure~\ref{fig:mscog051111} shows the offset between
$N_{best}$ and $N_{true}$ for each of the ions. The effects for
Si$^+$, Zn$^+$, and Mn$^+$ are expressed as upper limits (i.e.\ the offset
is at least as large as indicated) because the $N_{true}$ values derived
from the HIRES observations are conservative lower limits due to 
line saturation.   It is evident that the abundances for all of the
ions are systematically underestimated with offsets of 0.15\,dex to
at least 0.7\,dex.   The case of Zn$^+$ is particularly problematic.
The measurements consist of two
\ion{Zn}{2} transitions with $f\lambda$ values that
differ by 0.3\,dex. The observations are well fit by the MISC-COG analysis
yet $\N{Zn^+}_{best}$ is at least 0.5~dex too low.
The explanation is relatively simple:  the effective Doppler is so
large that these saturated profiles 
are treated as if they lie on the linear portion of the COG, i.e., 

\begin{equation}
N= \frac{W_\lambda}{\lambda} \; \frac{1.1 \sci{20} \cm{-2}}{f(\lambda/{\rm \AA})}
\label{eqn:weak}
\end{equation}
with $\lambda$ in Angstroms.  To summarize, the analysis is driven by
the strong \ion{Fe}{2} transitions to a large $b_{eff}$ value which leads
to a systematic underestimate of the column density of all the ions.
As worrisome, the underestimate varies from ion to ion and 
one cannot even calculate relative abundances to uncertainties of less
than 0.3\,dex.

\begin{figure}[ht]
\begin{center}
\includegraphics[width=3.6in]{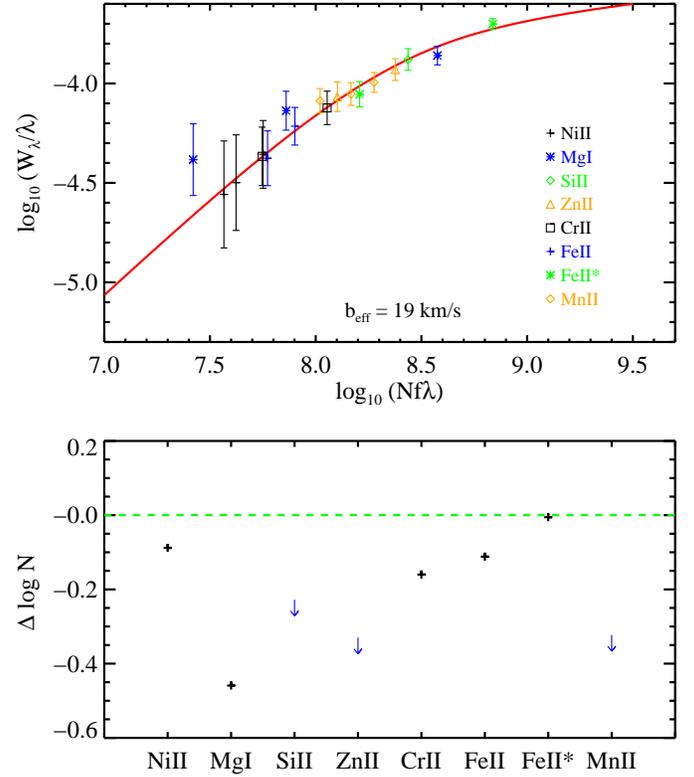}
\caption{Same as Figure~5 except we have removed
the strongest \ion{Fe}{2} transitions from the analysis.
The results are significantly improved due to the lower $b_{eff}$ value,
yet the column densities of \ion{Mg}{1},
\ion{Si}{2}, \ion{Zn}{2}, and \ion{Mn}{2} are still
significantly underestimated.
}
\label{fig:wkmscog051111}
\end{center}
\end{figure}

Not surprisingly, the results are improved if one eliminates the 
four strongest \ion{Fe}{2} transitions from the analysis 
(Figure~\ref{fig:wkmscog051111}).  We find, however, that we
still derive a $b_{eff}$ value that is too large and therefore 
significantly underestimate the column densities of several of the
ions.  If we only include those transitions which are unsaturated
at echelle resolution\footnote{This does not include 
Si$^+$ or Zn$^+$ whose lines have $\tau_0 > 5$.}  
(i.e.\ consider only those lines which are
truly on the linear portion of the COG), one finally achieves
reasonable results.   With low-resolution spectroscopy, however, one
does not have this luxury.
We are particularly worried that low-resolution spectroscopy
is generally limited to the analysis of transitions with
$\log (W_\lambda/\lambda) > -4$.

\begin{figure}[ht]
\begin{center}
\includegraphics[width=3.6in]{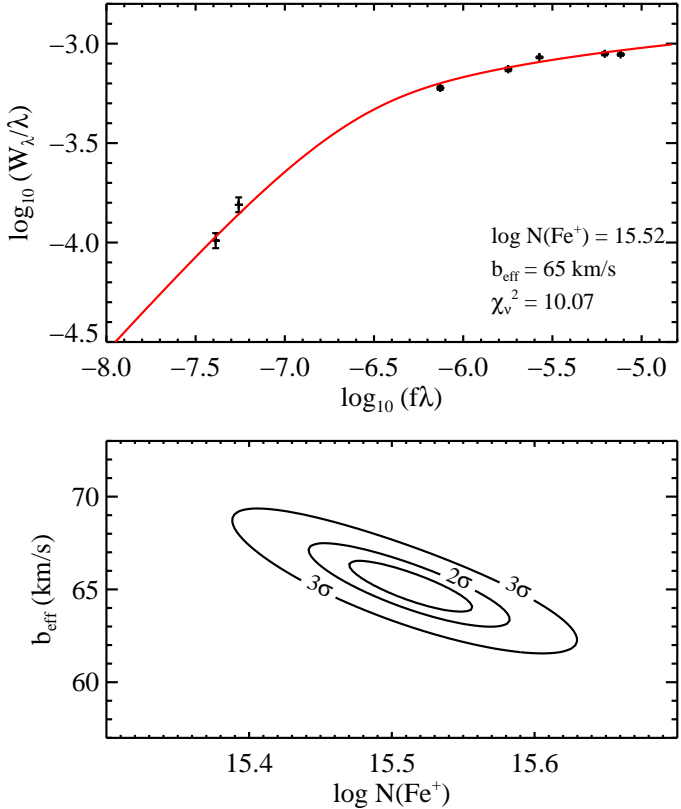}
\caption{Single-component COG analysis of the \ion{Fe}{2} transitions
for GRB~020813.
The upper panel
presents the $W_\lambda$ measurements and the model which minimizes
$\chi^2$.  The lower panel shows $\Delta \chi^2$ contours for the 
$b_{eff}$, $N$ parameter space (note the tiny dynamic range).  
}
\label{fig:sngcog020813}
\end{center}
\end{figure}

\section{GRB~020813: A High SNR, Low-Resolution Example from the Literature}
\label{sec:020813}

Let us now consider a low-resolution spectroscopic example from the
literature, GRB~020813.  \cite{bsc+03} obtained a high signal-to-noise,
low-resolution spectrum of this afterglow with the LRIS
spectrometer \citep{occ+95} in spectropolarimetery mode.  
\cite{bsc+03} reported equivalent width measurements for many
transitions associated with the ISM surrounding GRB~020813.  We present
in Figure~\ref{fig:sngcog020813} a single-component COG analysis
of the \ion{Fe}{2} transitions detected in their data.
Similar to GRB~051111 (Figure~\ref{fig:sngcog051111}),
the strong transitions drive the best-fit model to a large $b_{eff}$ value.
In contrast to GRB~051111, however, the weak Fe\,II~2249 and
Fe\,II~2260 transitions are well-modeled by this fit.  The error ellipse
for this model indicates a very precise measurement,
$\N{Fe^+} = 10^{15.52 \pm 0.03} \cm{-2}$, in good agreement with
the line-profile analysis of \cite{sf04}.  The correspondence
between the two analyses does not lend additional confidence;
for unresolved features, a single-component line-profile analysis
is identical to the COG method.

Because of the good visual agreement of the model (formally the
$\chi^2_\nu$ value is very poor) and the very small error ellipse,
one is led to believe that the column density has been precisely
determined.  This is an incorrect conclusion.  First, consider
the best-fit value $N_{best} = 10^{15.52} \cm{-2}$.  Adopting
$b_{eff} = 65 \mkms$, the line-profile of the \ion{Fe}{2}~2249
and 2260 transitions are optically thin, i.e.\ the transitions
are predicted to lie on the linear portion of the COG 
(Figure~\ref{fig:sngcog020813}).  In this case, one
can calculate $\N{Fe^+}$ from Equation~\ref{eqn:weak} and
find values of $10^{15.45} \cm{-2}$ and $10^{15.50} \cm{-2}$
for \ion{Fe}{2}~2249 and 2260 respectively.
While this result appears to be self-consistent, consider the
implication:  the rest equivalent widths of these two transitions
are very large, in fact $2.5\times$ larger than the partially
saturated \ion{Fe}{2} transitions for GRB~051111 
(Figure~\ref{fig:fe051111}).  To account for these large equivalent
widths and the best-fit column density, one must hypothesize a
line-profile consisting of a series of non-overlapping components 
that have large column density, but not too large 
($N \sim 10^{15} \cm{-2}$).  While this cannot be ruled out by
the observations, we consider this scenario to be highly unlikely
in light of the observed line-profiles of GRB sightlines
at high-resolution (e.g.\ Figure~\ref{fig:fe051111}).

\begin{figure}[ht]
\begin{center}
\includegraphics[width=3.6in]{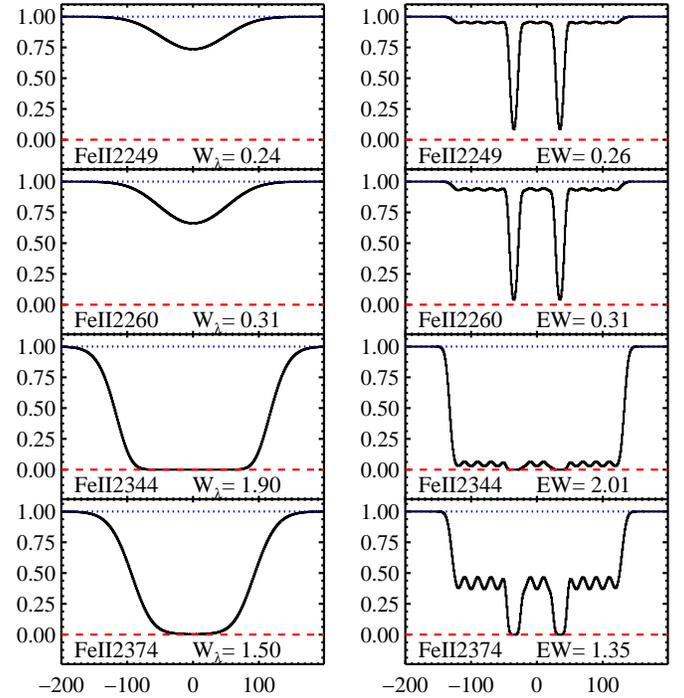}
\caption{Model profiles for the \ion{Fe}{2} transitions
of GRB~020813.  The LHS shows the single-component profile
corresponding to the best-fit model for Figure~\ref{fig:degeneracy}.
Note that the two \ion{Fe}{2} transitions are optically thin even
though they exhibit relatively large $W_\lambda$ values.
The RHS shows a toy model of the line-profile consisting of two
strong, narrow components and a series of weak features.
It presents a good match to the observed equivalent widths yet
has 100\% higher column density than the single-component model.
}
\label{fig:degeneracy}
\end{center}
\end{figure}

To emphasize this point, consider Figure~\ref{fig:degeneracy} where
we compare the single-component scenario with an alternate model.
The profiles on the left-hand side show several of
the observed \ion{Fe}{2} transitions predicted by the best-fit model.
The right-hand side shows a toy model consisting of two 
strong, narrow components (each with $\log N=15.4$ and $b=5 \mkms$)
and a series of weak components included to reproduce the
large equivalent width of the stronger \ion{Fe}{2} transitions.
This toy model provides a reasonable match to the observed
equivalent widths (in fact, a better fit to the strong transitions)
and implies a total Fe$^+$ column density that is two times larger
(0.3\,dex or $10\sigma$) than the single-component COG result.

We do admit, however, that it may be difficult to construct an
optical depth profile for GRB~020813 where the true Fe$^+$
column density is 10$\times$ larger than the single-component COG value.
And, we note that the low SNR, high-resolution VLT/UVES observations
of GRB~020813 \citep{fdl+05} do suggest that the optical depth
profile is not dominated by only a single strong component.
Nevertheless, we consider it impossible that the best-fit
COG value is not underestimating the $\N{Fe^+}$ value.  Furthermore,
we are certain that the statistical error estimate grossly underestimates
the systematic uncertainty associated with one's assumptions about
the optical depth profile.

\begin{figure}[ht]
\begin{center}
\includegraphics[height=3.6in,angle=90]{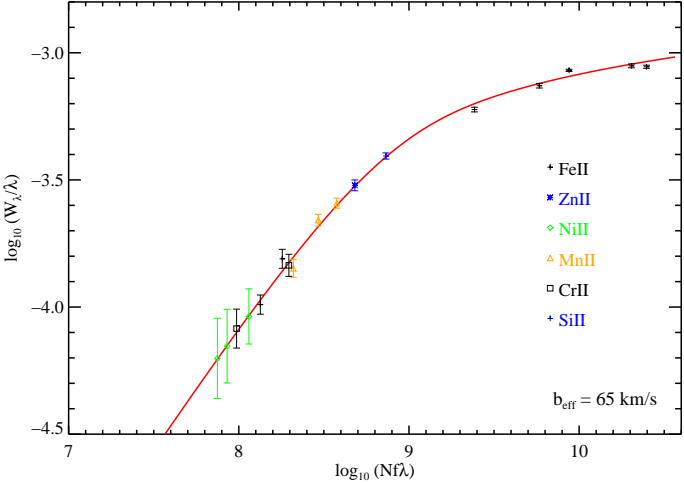}
\caption{
MISC-COG analysis of the ions in the ISM of GRB~020813.
Because of the large $b_{eff}$ values demanded by the heavily
saturated \ion{Fe}{2} transitions, the other transitions are
treated as if they lie on the linear portion of the curve-of-growth.
This leads to unrealistically low central values for the best-fit
column densities and overly optimistic error estimates.
}
\label{fig:mscog020813}
\end{center}
\end{figure}

Continuing with the treatment of GRB~020813, Figure~\ref{fig:mscog020813} 
presents the MISC-COG analysis for a series of the ions detected
in the \cite{bsc+03} spectrum.  Again, the model is visually
a good fit to the observations and we derive column densities
that are in good agreement with \cite{sf04}.  The issues that
exist for Fe$^+$ are accentuated for the Si$^+$ and Zn$^+$
results.  Here, the rest equivalent widths are 
$W_\lambda = 0.7$\AA\ and 0.6\AA\ for the \ion{Si}{2}~1808 and
\ion{Zn}{2}~2026 profiles, i.e.\ very large.
Yet, again, the best-fit column densities correspond to the 
optically thin limit because of the large $b_{eff}$ value.
It is nearly impossible to construct optically thin \ion{Si}{2}~1808 
and \ion{Zn}{2}~2026 line-profiles if one is restricted to 
physically realistic individual clouds with $b < 10\mkms$. 
Furthermore, {\it there is no observational constraint 
which sets an upper limit to the Si$^+$ and Zn$^+$ column densities}.

These are generic results for any MISC-COG analysis
that gives a very large $b_{eff}$ value:  the technique
models transitions with very large equivalent width measurements 
as being optically thin and adopts the lowest column density
conceivable.   The bottom line is that one is very likely
to underestimate the column densities of all of these ions.
And, the error estimates should reflect a one-sided 
distribution allowing only significantly larger values, i.e.\
lower limits.

\section{Conservative Suggestions for Equivalent Width Analysis}
\label{sec:proc}

Given the cautionary (pessimistic) tone of the previous
sections, one may draw the conclusion that no progress on chemical
abundances can
be made with low-resolution GRB spectroscopy.  To push forward in 
a productive yet cautious manner, we offer the following
suggestions.  

1. Assume that every line-profile is saturated and examine all
possible constraints to demonstrate otherwise. 
Absent additional constraints, the equivalent width measurements
establish only lower limits to the column densities.

2.  Consider independent analyses of the heavily saturated and the
weakest lines.  The former may give estimates for the abundances
of the ambient ISM along the sightline while the former could be
more relevant to the gas near the star-forming region.  
Unfortunately, one cannot separate the \ion{H}{1} gas along 
the sightline into these two phases so
the goals of such analysis would be to (i) prevent the strong
strong lines from biasing the total column density to low values 
and (ii) attempt to measure relative abundances in the two `phases'.
We reemphasize that by considering one COG
analysis for all the lines, the strong lines will drive the COG analysis 
to large $b_{eff}$ values which
forces the column densities of weaker (yet likely
saturated) transitions to their minimum value.
The end result is a systematic underestimate of all column densities.  
A corollary to this suggestion is that the results
for a single-component COG analysis 
with $b_{eff} > 20\mkms$ are highly suspect.

\begin{table}[ht]\footnotesize
\begin{center}
\caption{{\sc WEAK LINES FOR UPPER LIMITS\label{tab:wklin}}}
\begin{tabular}{lccccr}
\tableline
\tableline
Transition & $\lambda$ & $\log f$ & $N_{\tau_0=2}^a$
& $W_\lambda^b$ & [M/H]$^c$ \\
& (\AA) & & ($\cm{-2}$) & (m\AA) \\
\tableline
   SII 1250   & 1250.5840 &$ -2.2634$&15.29 & 82 &$-1.91$\\
    OI 1355   & 1355.5977 &$ -5.9066$&18.90 & 89 &$ 0.16$\\
   BII 1362   & 1362.4611 &$ -0.0057$&13.00 & 90 &$ 0.21$\\
  NiII 1467b  & 1467.7560 &$ -2.0044$&14.96 & 97 &$-1.29$\\
   PII 1532   & 1532.5330 &$ -2.1186$&15.06 &101 &$-0.47$\\
  GeII 1602   & 1602.4863 &$ -0.8428$&13.76 &106 &$ 0.13$\\
  FeII 1611   & 1611.2004 &$ -2.8665$&15.79 &106 &$-1.71$\\
  NiII 1703   & 1703.4050 &$ -2.2218$&15.12 &112 &$-1.13$\\
   MgI 1747   & 1747.7937 &$ -2.0419$&14.93 &115 &$-2.65$\\
  FeII 1901   & 1901.7729 &$ -3.9961$&16.84 &125 &$-0.66$\\
  CrII 2066   & 2066.1609 &$ -1.2882$&14.10 &136 &$-1.57$\\
  FeII 2249   & 2249.8767 &$ -2.7397$&15.51 &148 &$-1.99$\\
  SiII 2335   & 2335.1230 &$ -5.3716$&18.13 &154 &$ 0.57$\\
  TiII 3230   & 3230.1311 &$ -1.1630$&13.78 &213 &$-1.16$\\
\tableline
\end{tabular}
\end{center}
\tablenotetext{a}{Column density corresponding to a line-profile with peak optical
 depth \\
$\tau_0 = 2$ and Doppler parameter $b=10\mkms$.}
\tablenotetext{b}{Rest equivalent width for the column density in column 4.}
\tablenotetext{c}{Metallicity assuming the column density in column 4 and $\log \mnhi = 22$.}
\end{table}

3. High precision measurements, absolute or relative, will be very
difficult to achieve.
Meaningful constraints that require better than 0.5\,dex precision  
(e.g.\ nucleosynthetic patterns) may not be attainable.
One may, however, be able to distinguish between
very high and low dust-to-gas ratios.

4. Aim for the highest spectral resolution at the cost of
high signal-to-noise.   The only convincing argument that a
line is unsaturated is to reconstruct the optical depth profile.

5.  Place upper limits on the metallicity using very weak transitions
and/or trace elements.  Table~\ref{tab:wklin} provides a list of 
weak transitions for abundant elements (e.g.\ Fe, Si, O)
and stronger transitions of trace elements  (e.g.\ Ge, B).
The Table also lists the column density corresponding to a
line-profile which is modestly saturated $(\tau_0 = 2)$ for
a Doppler parameter $b=8\mkms$.  If one can establish an
upper limit (ideally $3\sigma$ including continuum uncertainty)
to the equivalent width which is below the corresponding 
$W_\lambda$ value (Column~5) then one places an upper limit
to the metallicity listed in Column~6, assuming solar relative
abundances and ignoring differential depletion.  In general,
this will require observations with rest equivalent width
error $\sigma(W_\lambda) < 30$m\AA.

\clearpage

\begin{table*}[ht]\footnotesize
\begin{center}
\caption{{\sc SUMMARY TABLE OF ABUNDANCE MEASUREMENTS IN LOW-RESOLUTION GRB SPECTRA\label{tab:summ}}}
\begin{tabular}{lccccclcccc}
\tableline
\tableline
GRB & $z$ & $\log \mnhi$ & [M/H]$_{lit}^a$
& Ref & Instr & Line$_{weak}^b$ & 
$W_\lambda$(\AA) & [M/H]$_{min}^c$ & [M/H]$_{max}^d$ \\
\tableline
990123 & 1.600 & 22              & $-0.7^{+0.05}_{-0.05}$  & 1,2   & Keck/LRIS    & \ion{Zn}{2}~2026 & $0.485 \pm 0.038$ & $-0.95$  & $-$ \\
000926 & 2.038 & $21.3 \pm 0.25$ & $-0.15^{+0.05}_{-0.05}$ & 2,3,4 & Keck/ESI     & \ion{Zn}{2}~2026 & $0.90 \pm 0.1$    & $-0.25$  & $-$ \\
010222 & 1.477 & 22              & $-1.0^{+0.07}_{-0.07}$  & 5,2   & Keck/ESI     & \ion{Zn}{2}~2026 & $0.91 \pm 0.1$    & $-1.0 $  & $-$ \\
011211 & 2.142 & $20.4 \pm 0.2$  & $-0.9^{+0.6}_{-0.4}$    & 6,2   & VLT/FORS2    & \ion{Zn}{2}~2026 & $0.91 \pm 0.1$    & $-1.0 $  & $-$ \\
020813 & 1.255 & 22              & $-1.1^{+0.06}_{-0.06}$  & 7,2   & Keck/LRIS    & \ion{Zn}{2}~2026 & $0.61 \pm 0.1$    & $-1.2 $  & $-$ \\
030323 & 3.372 & $21.9 \pm 0.07$ & $-1.3^{+0.20}_{-0.20}$  & 8     & VLT/FORS2    & \ion{Zn}{2}~2026 & $0.82 \pm 0.09$   & $-0.9 $  & $+0.5$   \\
050401 & 2.899 & $22.5 \pm 0.3$  & $-0.8^{+0.4}_{-0.4}$  & 9     & VLT/FORS2    & \ion{Zn}{2}~2062 &                   & $-1.0 $  & $-$ \\
050505 & 4.275 & $22.05\pm 0.1$  & $-1.2                $  & 10    & Keck/LRIS    & \ion{Fe}{2}~1611 & $0.54 \pm 0.05$   & $-1.0 $  & $-$ \\
050904 & 6.296 & $21.3 \pm 0.2$  & $-1                  $  & 11    & Subaru/FOCAS & \ion{S}{2}~1253  & $0.5  \pm 0.15$   & $-1.4 $  & $-$ \\
060206 & 4.048 & $20.85\pm 0.1$  & $-0.84^{+0.1}_{-0.1}     $  & 12    & WHT/ISIS     & \ion{S}{2}~1250  &                   & $-1.0 $  & $-0.5$   \\
\tableline
\end{tabular}
\end{center}
\tablenotetext{a}{Metallicity reported in the literature.  For those sightlines without
\nhi\ measurements, we adopt a reasonably conservative value of 10$^{22} \cm{-2}$.  Note that the errors
reported reflect only the uncertainy in the reported metal abundance to highlight the overly
optimistic results.}
\tablenotetext{b}{Most constraining transition for a metallicity measurement, generally derived from a weak transition.  Note that we have not corrected
for \ion{Mg}{1} blending with the \ion{Zn}{2}~2026 transition.   This could reduce the metallicity limit by 
as much as 0.2\,dex.}
\tablenotetext{c}{Minimum metallicity derived from assuming the transition in Column~8 is 
optically thin.}
\tablenotetext{d}{Maximum metallicity based on the non-detection of a transition or the
accurate measurement of an unsaturated transition.}
\tablerefs{
1: \cite{sff03}; 
2: \cite{kdo+99}; 
3: \cite{fgm+01};
4: \cite{cgh+03};
5: \cite{mhk+02};
6: \cite{vsf+06};
7: \cite{bsc+03};
8: \cite{vel+04};
9: \cite{watson050401};
10: \cite{bpck+05};
11: \cite{kka+06};
12: \cite{fynbo060206}
}
\end{table*}

\section{Concluding Remarks}

Before concluding, we wish to briefly comment on the abundance
measurements for the ISM of GRB presented in the literature.
Table~\ref{tab:summ} summarizes the results to date.
The last two columns give the minimum and maximum metallicity values
allowed by the data as reported in the 
literature\footnote{We will ignore (for the time being) 
the fact that Zn is a
trace element and that a large Zn/H ratio does not require a
high gas metallicity.}.  One notes that solar (and even super-solar)
abundances are allowed in nearly every case.  
It is evident from Table~\ref{tab:summ} that 
the principal conclusion of \cite{sff03} -- GRB sightlines 
have extreme metal column densities -- may be a remarkable understatement. 
We also caution that the common conclusion that GRB occur in a metal-poor 
ISM ([M/H]~$\leq -1$) may need serious revision.

It is unfortunate that there are few transitions for mild or non-refractory
elements that are weak enough to place meaningful upper limits 
to the abundances;
the Zn and Si abundances, for example, are compromised by their 
relatively strong transitions. 
We expect, however, that an analysis of the lines listed in 
Table~\ref{tab:wklin} may improve the constraints \citep{pro_dataI06}
and provide more realistic metallicity measurements for the gas
within GRB host galaxies.

\acknowledgments

The authors wish to recognize and acknowledge the very significant
cultural role and reverence that the summit of Mauna Kea has always
had within the indigenous Hawaiian community.  We are most fortunate
to have the opportunity to conduct observations from this mountain.
J.X.P. thanks E. Jenkins, H.-W. Chen, and J.S. Bloom for valuable
discussions and comments on a draft of this manuscript.
J.X.P. is partially supported by NASA/Swift grant NNG05GF55G.



\end{document}